\documentclass[12pt]{article}
\usepackage[dvips]{color}
\usepackage{epsfig}
\usepackage{amsmath}
\usepackage{graphicx}

\begin{document}
\title{\bf Gravitational lensing of STU black holes}
\author{{H. Saadat\thanks{Email: hsaadat2002@yahoo.com}}\\
{\small {\em  Department of  physics, Shiraz Branch, Islamic Azad
University, Shiraz, Iran}}\\
{\small {\em P. O. Box 71555-477, Iran}}} \maketitle
\begin{abstract}
In this paper we study gravitational lensing by STU black holes. We
considered extremal limit of two special cases of zero-charged and
one-charged black holes, and obtain the deflection
angle. \\\\
\noindent {\bf Keywords:} Gravitational lensing; STU black hole.
\end{abstract}
\section{Introduction}
Strong and weak gravitational lensing is one of the interesting
subjects of recent studies in theoretical physics, cosmology and
astrophysics [1–7]. The investigation of gravitational lensing is
powerful method to probe the extra dimensions. In that case the
gravitational lensing by squashed Kaluza-Klein black holes studied
in Refs. [8, 9] and developed to charged squashed Kaluza-Klein black
hole by the Ref. [10].\\
Also there are interesting relation between the gravitational
lensing and dark energy. So in the Ref. [11] the effect of phantom
scalar field (as a candidate for dark energy) on the gravitational
lensing has been studied. In the similar way one can study effect of
Chaplygin gas [12, 13] on the gravitational lensing.\\
Now, aim of this paper is studying the gravitational lensing by
so-called STU black hole [14-20]. The STU black hole exist in
special case of $D=5$, $\mathcal{N}=2$ gauged supergravity theory
which is dual to the $\mathcal{N}=4$ SYM theory with finite chemical
potential. In this background, generally there are three electric
charges. In this paper we assume that two of them will be zero and
discussed about one-charged black hole.
\section{Gravitational lensing}
For a given generic black hole background of the form,
\begin{equation}\label{s1}
ds^2=-A(r)dt^2+B(r)dr^2+C(r)d\Omega^{2},
\end{equation}
it is useful to define the following quantity [21],
\begin{equation}\label{s2}
X=u\sqrt{\frac{B(r)}{C(r)[\frac{C(r)}{A(r)}-u^{2}]}}.
\end{equation}
Therefore, the deflection angle is given by,
\begin{equation}\label{s3}
\alpha=-\pi+2\int_{r_{m}}^{\infty}{X}dr,
\end{equation}
where,
\begin{equation}\label{s4}
u=\sqrt{\frac{C(r_{m})}{A(r_{m})}},
\end{equation}
is the impact parameter and $r_{m}$ is closest distance between
traveling photon and the black hole. Now, if we assume a source at
coordinates ($r_{S}$, $\phi_{S}$) and an observer at ($r_{O}$,
$\phi_{O}$), then the total azimuthal shift experienced by the
photon is given by,
\begin{equation}\label{s5}
\Phi_{-}=\int_{r_{m}}^{r_{O}}{X}dr+ \int_{r_{m}}^{r_{S}}{X}dr.
\end{equation}
Also it is possible that the photon directly goes from the source to
the observer far from the black hole. In that case the azimuthal
shift is given by,
\begin{equation}\label{s6}
\Phi_{+}=\int_{r_{S}}^{r_{O}}{X}dr.
\end{equation}
Therefore, lens equation will be $\Phi=\Phi_{-}$ if
$\triangle\phi>\Phi_{\pm}$ and  $\Phi=\Phi_{+}$ if
$\triangle\phi\leq\Phi_{\pm}$, where
\begin{equation}\label{s7}
\triangle\phi=\phi_{O}-\phi_{S}+2n\pi.
\end{equation}
Also, the angle $\theta$ where the observer detects the photon is
given by the following expression,
\begin{equation}\label{s8}
\theta=\sin^{-1}\left(u\sqrt{\frac{A(r_{O})}{C(r_{O})}}\right).
\end{equation}
\section{STU black hole}
STU black hole is described by the following solution,
\begin{equation}\label{s9}
ds^{2}=-\frac{f}{{\mathcal{H}}^{\frac{2}{3}}}dt^{2}
+{\mathcal{H}}^{\frac{1}{3}}(\frac{dr^{2}}{f}+\frac{r^{2}}{R^{2}}d\Omega^{2}),
\end{equation}
where,
\begin{eqnarray}\label{s10}
f&=&1-\frac{\mu}{r^{2}}+\frac{r^{2}}{R^{2}}{\mathcal{H}},\nonumber\\
{\mathcal{H}}&=&\prod_{i=1}^{3} H_{i},\nonumber\\
H_{i}&=&1+\frac{q_{i}}{r^{2}}, \hspace{10mm} i=1, 2, 3,\nonumber\\
A_{t}^{i}&=&\sqrt{\frac{q_{i}+\mu}{q_{i}}}(1-H_{i}^{-1}),
\end{eqnarray}
and $R$ is the constant AdS radius which relates to the coupling
constant via $R=1/g$ (also, coupling constant relates to the
cosmological constant via $\Lambda=-6g^2$), and $r$ is the radial
coordinate along the black hole. The black hole horizon specified by
$r=r_{h}$ which is obtained from $f=0$. Also $\mu$ is called
non-extremality parameter. So, for the extremal limit one can assume
$\mu=0$. The Hawking temperature of STU black hole is given by,
\begin{equation}\label{s11}
T=\frac{r_{h}}{2\pi
R^{2}}\frac{2+\frac{1}{r_{h}^{2}}\sum_{i=1}^{3}{q_{i}}-\frac{1}{r_{h}^{6}}\prod_{i=1}^{3}{q_{i}}}{{\sqrt{\prod_{i=1}^{3}(1+\frac{q_{i}}{r_{h}^{2}})}}}.
\end{equation}
So, in the case of $q_{i}=0$ we get,
\begin{equation}\label{s12}
r_{h}=\pi R^{2}T,
\end{equation}
and for the case of one-charged black hole ($q_{1}=q$,
$q_{2}=q_{3}=0$) one can obtain,
\begin{equation}\label{s13}
r_{h}=\frac{1}{2}\sqrt{2\pi^{2}
R^{4}T^{2}\left(1+\sqrt{1+\frac{2q}{\pi^{2} R^{4}T^{2}}}\right)-2q},
\end{equation}
which is reduced to the equation (12) for $q=0$.\\
It should be pointed out that the solution (10) written for the
spherical space with curvature $k=1$. It is also possible to
consider the case of flat space with $k=0$, so one can write [22],
\begin{equation}\label{s14}
f=-\frac{\mu}{r^{2}}+\frac{r^{2}}{R^{2}}{\mathcal{H}}.
\end{equation}
\section{Uncharged black hole}
First of all we consider very special case of zero-charged black
hole with $q=0$. Then using line element (9) in the equations (2),
(3) and (4) gives the following deflection angle,
\begin{eqnarray}\label{s15}
\alpha&=&-\pi+2R(\ln2+\ln R)\nonumber\\
&+&2\,R \left( \ln \left( {\frac {r_{m}}{\sqrt {{R}^{2}+r_{m}^{2}}}}
\right) +\ln \left( \sqrt {{\frac
{r_{m}^{4}(1-R^{2})+r_{m}^{2}R^{2}}{{R}^{2}+r_{m}^{2}}}} \right) -
\ln \left( r_{m} \right) \right).
\end{eqnarray}
Then, by using the equation (8) one can obtain the angle $\theta$ as
the following,
\begin{equation}\label{s16}
\theta=\sin^{-1}\left(\frac{r_{m}^{2}\sqrt{R^{2}+r_{O}^{2}}}{r_{O}^{2}\sqrt{R^{2}+r_{m}^{2}}}\right)
\end{equation}
In the Fig. 1 we draw the deflection angle in terms of closest
distance of a photon with the black hole. Interesting assumption is
that the photon reach the near horizon ($r_{m}\approx r_{h}$), in
that case one can obtain the deflection angle in terms of the black
hole temperature,
\begin{equation}\label{s17}
\alpha=\pi -R\left(\ln  \left( 2 \right) +\ln  \left( \pi \right)
+2\ln
 \left( R \right) +\ln  \left( T \right) -\ln  \left( 1+{\pi }^{2}{R
}^{2}{T}^{2} \right)\right),
\end{equation}
which is illustrated in The Fig. 2. It shows that the black hole
temperature increases the deflection angle.

\begin{figure}[th]
\begin{center}
\includegraphics[scale=.3]{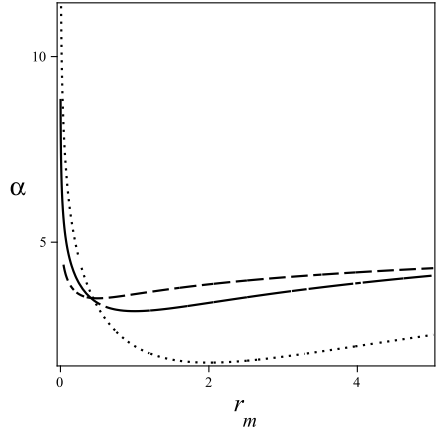}
\caption{Deflection angle versus $r_{m}$ for $R=0.5$ (dashed line),
$R=1$ (solid line) and $R=2$ (dotted line). }
\end{center}
\end{figure}

\begin{figure}[th]
\begin{center}
\includegraphics[scale=.3]{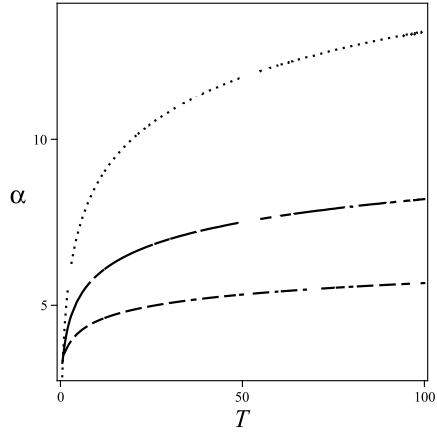}
\caption{Deflection angle versus $T$ for $R=0.5$ (dashed line),
$R=1$ (solid line) and $R=2$ (dotted line). }
\end{center}
\end{figure}

\section{One-charged black hole}
In order to find the effect of black hole charge on the deflection
angle, we consider the simplest case of one-charged black hole. In
that case we set $q_{1}=q$ and $q_{2}=q_{3}=0$. Therefore we find,
\begin{equation}\label{s18}
\alpha=-\pi+\frac{2R\sqrt{r_{m}^{2}+q}}{r_{m}}\left(\ln2+2\ln
R+\ln{r_{m}}-\ln{r_{m}^{2}+R^{2}+q}\right).
\end{equation}
In the Fig. 3 we find that the black hole charge increased the
deflection angle. Then the angle $\theta$ obtained as the following,
\begin{equation}\label{s19}
\theta=\sin^{-1}\left(\frac{\sqrt{r_{m}^{2}+q}\sqrt{R^{2}+r_{O}^{2}+q}}{\sqrt{r_{O}^{2}+q}\sqrt{R^{2}+r_{m}^{2}+q}}\right)
\end{equation}

\begin{figure}[th]
\begin{center}
\includegraphics[scale=.3]{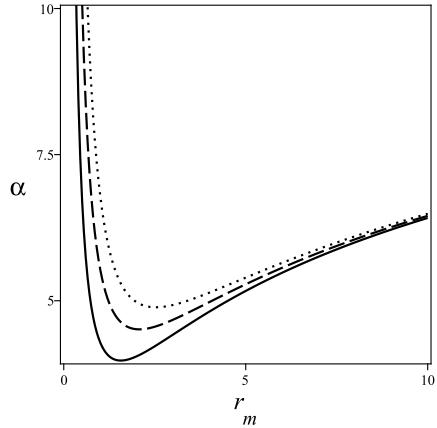}
\caption{Deflection angle versus $r_{m}$ with $R=1$ for $q=1$(solid
line), $q=2$ (dashed line) and $q=3$ (dotted line).}
\end{center}
\end{figure}

\section{Conclusion}
In this paper we considered one-charged black hole of STU model at
extremal limit to find the effect of the black hole charge on the
gravitational lensing. In that case we found that the black hole
charge increased the deflection angle. This work will be extended to
the three-charged non-extremal black hole, however it is expected
that the black hole charge increased the deflection angle.\\
Also it is interesting to study gravitational lensing by other
important black holes. For instant there are a class of two
dimensional extremal black holes [23], BTZ black holes [24], or
Ho\v{r}ava-Lifshitz black hole [25]. Also there are new works, where
some black holes extend to the hyperscaling version [26, 27], which
are interesting subjects to study gravitational lensing.


\begin{thebibliography}{11}
\bibitem{P1}
E.F. Eiroa, G.E. Romero and D.F. Torres, Phys. Rev. D 65 (2002)
024010
\bibitem{P2}
M. Sereno, Phys. Rev. D 67 (2003) 064007
\bibitem{P3}
A. Bhadra, Phys. Rev. D 67 (2003) 103009
\bibitem{P4}
V. Bozza, Phys. Rev. D 66 (2002) 103001
\bibitem{P5}
V. Bozza and L. Mancini, Phys. Rev. D 69 (2004) 063004
\bibitem{P6}
S. Frittelli, T.P. Kling and E.T. Newman, Phys. Rev. D 61 (2000)
064021
\bibitem{P7}
K.S. Virbhadra and G.F.R. Ellis, Phys. Rev. D 62 (2000) 084003
\bibitem{P8}
Y. Liu, S. Chen and J. Jing, Phys. Rev. D 81 (2010) 124017
\bibitem{P9}
S. Chen, Y. Liu and J. Jing, Phys. Rev. D 83 (2011) 124019
\bibitem{P10}
J. Sadeghi, A. Banijamali, H. Vaez, "Strong Gravitational Lensing in
a Charged Squashed Kaluza- Klein Black hole", [arXiv:1205.0805
[gr-qc]]
\bibitem{P11}
G.N. Gyulchev, I.Z. Stefanov, "Gravitational Lensing by Phantom
Black holes", Phys. Rev. D 87 (2013) 063005 [arXiv:1211.3458
[gr-qc]]
\bibitem{P12}
B. Pourhassan, "Viscous Modified Cosmic Chaplygin Gas Cosmology",
IJMPD (2013) [arXiv:1301.2788 [gr-qc]]
\bibitem{P13}
H. Saadat, B. Pourhassan, "Effect of Varying Bulk Viscosity on
Generalized Chaplygin Gas", [arXiv:1305.6054 [gr-qc]]
\bibitem{P14}
J. Sadeghi, B. Pourhassan, "Drag Force of Moving Quark at The N=2
Supergravity", JHEP 0812 (2008) 026, [arXiv:0809.2668 [hep-th]]
\bibitem{P15}
J. Sadeghi, et al., "Drag Force of Moving Quark in STU Background",
Eur.Phys.J.C61 (2009) 527, [arXiv:0901.0217 [hep-th]]
\bibitem{P16}
J. Sadeghi, M. R. Setare, B. Pourhassan, "Drag force with different
charges in STU background and AdS/CFT", J. Phys. G: Nucl. Part.
Phys. 36 (2009) 115005 [arXiv:0905.1466 [hep-th]]
\bibitem{P17}
J. Sadeghi, B.Pourhassan, A. Chatrabhuti, "$AdS_{5}$ black hole at
N=2 supergravity", Indian Journal of Physics 87 (2013) 691
[arXiv:0912.1423 [hep-th]]
\bibitem{P18}
K. B. Fadafan, B. Pourhassan, J. Sadeghi, "Calculating the
jet-quenching parameter in STU background", Eur. Phys. J. C 71
(2011) 1785, [arXiv:1005.1368 [hep-th]]
\bibitem{P19}
J. Sadeghi, B. Pourhassan, A. R. Amani, "The effect of higher
derivative correction on $\eta/s$ and conductivities in STU model",
Int. J. Theor. Phys. 52 (2013) 42, [arXiv:1011.2291 [hep-th]]
\bibitem{P20}
J. Sadeghi, B. Pourhassan, "STU/QCD Correspondence",
[arXiv:1205.4254 [hep-th]]
\bibitem{P21}
V. Bozza, "Gravitational Lensing by Black Holes", [arXiv:0911.2187
[gr-qc]]
\bibitem{P22}
H. Saadat, "R-Charged Black Hole and Holographic Superfluid", Int.
J. Theor. Phys. 51 (2012) 3471
\bibitem{P23}
J. Sadeghi, M. R. Setare, B.Pourhassan, "Entropy of Extremal Black
Holes in Two Dimension", Acta Phys. Polon. B40 (2009) 251,
[arXiv:0707.0420 [hep-th]]
\bibitem{P24}
J. Sadeghi et al., "Cosmic String in the BTZ Black Hole Background
with Time-Dependant Tension", Phys. Lett. B703 (2011) 14,
[arXiv:0903.0292 [hep-th]]
\bibitem{P25}
J. Sadeghi, B.Pourhassan, "Particle acceleration in Horava-Lifshitz
black holes" Eur. Phys. J. C 72 (2012) 1984, [arXiv:1108.4530
[hep-th]]
\bibitem{P26}
J. Sadeghi, B. Pourhassan, A. Asadi, "Thermodynamics of string black
hole with hyperscaling violation", [arXiv:1209.1235 [hep-th]]
\bibitem{P27}
J. Sadeghi, B. Pourhassan, F. Pourasadollah, "Thermodynamics of
Schrödinger black holes with hyperscaling violation", Physics
Letters B 720 (2013) 244 [arXiv:1209.1874 [hep-th]]
\end{thebibliography}
\end{document}